\documentclass[a4paper,10pt,twoside]{cpc-hepnp}

\usepackage{multicol}
\usepackage{graphicx}
\usepackage{booktabs}
\usepackage{amssymb,bm,mathrsfs,bbm,amscd}
\usepackage[tbtags]{amsmath}
\usepackage{lastpage}

\begin{document}



\newcommand{\be}{\begin{equation}}
\newcommand{\ee}{\end{equation}}
\newcommand{\bea}{\begin{eqnarray}}
\newcommand{\eea}{\end{eqnarray}}
\newcommand{\bef}{\begin{figure}}
\newcommand{\eef}{\end{figure}}
\newcommand{\bce}{\begin{center}}
\newcommand{\ece}{\end{center}}

\newcommand{\bfn}{{\bm {n}}}
\newcommand{\bfJ}{{\bm {J}}}

\newcommand{\pion}{{\pi}}
\newcommand{\kaon}{{K}}
\newcommand{\kaonb}{{\bar{K}}}
\newcommand{\kaonS}{{K^*}}
\newcommand{\kaonSb}{{\bar{K}^*}}
\newcommand{\D}{{D}}
\newcommand{\Db}{{\bar{D}}}
\newcommand{\Ds}{{D_s}}
\newcommand{\Dsb}{{\bar{D}_s}}
\newcommand{\DS}{{D^*}}
\newcommand{\DSb}{{\bar{D}^*}}
\newcommand{\DsS}{{D_s^*}}
\newcommand{\DsSb}{{\bar{D}_s^*}}
\newcommand{\etap}{{\eta^\prime}}
\newcommand{\etac}{{\eta_c}}
\newcommand{\Jpsi}{{\psi}}
\newcommand{\Nucleon}{{N}}
\newcommand{\Lambdac}{{\Lambda_c}}
\newcommand{\SigmaS}{{\Sigma^*}}
\newcommand{\Sigmac}{{\Sigma_c}}
\newcommand{\SigmacS}{{\Sigma_c^*}}
\newcommand{\Cascada}{{\Xi}}
\newcommand{\Cascadac}{{\Xi_c}}
\newcommand{\CascadaS}{{\Xi^*}}
\newcommand{\CascadacS}{{\Xi_c^*}}
\newcommand{\Cascadacc}{{\Xi_{cc}}}
\newcommand{\CascadaccS}{{\Xi_{cc}^*}}
\newcommand{\Cascadacp}{{\Xi_c^\prime}}

\newcommand{\Omegac}{{\Omega_c}}
\newcommand{\Omegacc}{{\Omega_{cc}}}
\newcommand{\OmegacS}{{\Omega_c^*}}
\newcommand{\OmegaccS}{{\Omega_{cc}^*}}
\newcommand{\Omegaccc}{{\Omega_{ccc}}}

\newcommand{\Qur}{{Q^\dagger_{u\uparrow}}}
\newcommand{\Quj}{{Q^\dagger_{u\downarrow}}}
\newcommand{\Qdr}{{Q^\dagger_{d\uparrow}}}
\newcommand{\Qdj}{{Q^\dagger_{d\downarrow}}}
\newcommand{\Qsr}{{Q^\dagger_{s\uparrow}}}
\newcommand{\Qsj}{{Q^\dagger_{s\downarrow}}}
\newcommand{\Qcr}{{Q^\dagger_{c\uparrow}}}
\newcommand{\Qcj}{{Q^\dagger_{c\downarrow}}}

\newcommand{\Qubr}{{Q^\dagger_{\bar{u}\uparrow}}}
\newcommand{\Qubj}{{Q^\dagger_{\bar{u}\downarrow}}}
\newcommand{\Qdbr}{{Q^\dagger_{\bar{d}\uparrow}}}
\newcommand{\Qdbj}{{Q^\dagger_{\bar{d}\downarrow}}}
\newcommand{\Qsbr}{{Q^\dagger_{\bar{s}\uparrow}}}
\newcommand{\Qsbj}{{Q^\dagger_{\bar{s}\downarrow}}}
\newcommand{\Qcbr}{{Q^\dagger_{\bar{c}\uparrow}}}
\newcommand{\Qcbj}{{Q^\dagger_{\bar{c}\downarrow}}}

\newcommand{\uno}{{\bf 1}}
\newcommand{\dos}{{\bf 2}}
\newcommand{\tres}{{\bf 3}}
\newcommand{\cuatro}{{\bf 4}}
\newcommand{\ocho}{{\bf 8}}
\newcommand{\diez}{{\bf 10}}
\newcommand{\quince}{{\bf 15}}
\newcommand{\dieciseis}{{\bf 16}}
\newcommand{\veinte}{{\bf 20}}
\newcommand{\veintisiete}{{\bf 27}}
\newcommand{\treintaycinco}{{\bf 35}}
\newcommand{\cincuentayseis}{{\bf 56}}
\newcommand{\sesentaytres}{{\bf 63}}
\newcommand{\cientoveinte}{{\bf 120}}
\newcommand{\cientosesentayocho}{{\bf 168}}
\newcommand{\setecientosveinte}{{\bf 720}}
\newcommand{\novecientoscuarentaycinco}{{\bf 945}}
\newcommand{\mildoscientostreintaydos}{{\bf 1232}}
\newcommand{\dosmilquinientosveinte}{{\bf 2520}}
\newcommand{\cuatromilsetecientoscicuentaydos}{{\bf 4752}}
\newcommand{\trecemilcientocuatro}{{\bf 13104}}

\title{Charmed baryon resonances \\ with heavy-quark symmetry}

\author{%
      L. Tolos$^{1;1)}$\email{tolos@kvi.nl}%
\quad C. Garcia-Recio$^{2}$
\quad V. K. Magas $^{3}$
\quad T. Mizutani $^{4}$
\quad J. Nieves $^{5}$
\quad A. Ramos $^{3}$
\quad L. L. Salcedo $^{2}$
}
\maketitle

\address{%
1~Theory Group, KVI, University of Groningen, Zernikelaan 25, 9747 AA
Groningen, The Netherlands\\
2~Departamento de F{\'\i}sica At\'omica, Molecular y Nuclear, Universidad de Granada,
 E-18071 Granada, Spain\\
3~Departament d'Estructura i
  Constituents de la Mat\`eria and Institut de Ci\`encies del Cosmos, \\ Universitat de Barcelona, E-08028 Barcelona, Spain\\
4~Department of Physics, Virginia
  Polytechnic Institute and State University, Blacksburg, VA 24061, USA\\
5~Instituto de F{\'\i}sica Corpuscular (centro mixto CSIC-UV),\\
Institutos de Investigaci\'on de Paterna, Aptdo. 22085, E-46071 Valencia, Spain\\
}

\begin{abstract}
We study charmed baryon resonances that are generated dynamically from a coupled-channel unitary approach that implements heavy-quark symmetry. Some states can already be identified with experimental observations, such as $\Lambda_c(2595)$, $\Lambda_c(2660)$, $\Sigma_c(2902)$ or $\Lambda_c(2941)$, while others need a compilation of more experimental data as well as an extension of the model to include higher order contributions.  We also compare our model to previous SU(4) schemes.
\end{abstract}



\begin{multicols}{2}

\section{Introduction}

Charmed baryon resonances have received recently a lot of attention in connection with the discovery of some new states by the CLEO, Belle and BABAR collaborations \cite{cleo,belle,babar,aubert}. In fact, one of the challenges in hadron physics over the last years is to establish whether a resonance has the usual $q \bar q$ or $qqq$ structure, or it better qualifies as being dynamically generated via unitarized meson-baryon scattering processes. The extension to the charm sector of the unitarized meson-baryon method in coupled channels was initially attempted in Ref.~\cite{tolos-schaffne-mishra}, where the free space amplitudes were constructed from a set of separable coupled-channel interactions obtained from chirally motivated lagrangians upon replacing the $s$ quark by the $c$ quark. A different approach resulting from the scattering of Goldstone bosons off the ground state $1/2^+$ charmed baryons was  pursued in \cite{kolo}, but the substantial improvement in constructing the meson-baryon interaction in the charm sector came from exploiting the universal vector-meson coupling hypothesis to
break the SU(4) symmetry \cite{hofmann}. The $t$-channel exchange of vector mesons (TVME) between pseudoscalar mesons and baryons preserved chiral symmetry in the light meson sector keeping the Weinberg-Tomozawa (WT) type of interaction. An extension to $d$-wave $J=3/2^-$ resonances was developed in \cite{hofmann2}, while some modifications over the model of Ref.~\cite{hofmann} were implemented in Ref.~\cite{angels-mizutani}, both in the kernel and in the renormalization scheme. More recently, there have been attempts to construct the $DN$ and $\bar DN$ interaction by incorporating the charm degree of freedom in the SU(3) meson-exchange model of the
J\"ulich group \cite{haide1}.
 
However, those SU(4) TVME inspired models are not consistent with heavy-quark symmetry (HQS), which is a proper QCD spin-flavor symmetry that appears when the quark masses, such as the charm mass, become
larger than the typical confinement scale. As a consequence of this symmetry, the spin interactions vanish for infinitely massive quarks. Thus, heavy hadrons come in doublets (if the spin of the light degrees of freedom is not zero), which are degenerated in the infinite quark-mass limit. First attempts were done in the strange sector by means of a scheme that starts from a SU(6) spin-flavor symmetry Lagrangian and that incorporates some symmetry breaking corrections determined by physical masses and meson decay constants~\cite{Garcia-Recio:2005hy,GarciaRecio:2006wb,Toki:2007ab}.  The corresponding Bethe-Salpeter equation reproduces the previous SU(3)-flavor WT results for the lowest-lying $s$- and $d$-wave, negative parity baryon resonances and gives new information
on more massive states, as for example the $\Lambda(1800)$ or $\Lambda(2325)$ resonances.

In this paper, we extend this scheme to four flavors, incorporating the charm degree of freedom ~\cite{magas}. This approach automatically incorporates HQS in the charm sector improving in this respect on the  SU(4) TVME models. We focus on non-strange single charmed resonances close to their relevant thresholds and compare our results to the previous SU(4) models. Of several novelties, we find that  the dynamics of the lowest lying resonance
$\Lambda_c(2595)$ is completely dominated by the $DN$ channel in the SU(4) TVME model \cite{hofmann}, while it turns out be largely a $D^*N$ state within the SU(8) scheme \cite{magas}.

\section{SU(8) extension of the WT meson-baryon lagrangian}

The WT Lagrangian is not only SU(3) symmetric but also chiral invariant. Symbolically, up to
an overall constant, the WT interaction is \be {\cal L_{\rm WT}}= {\rm
Tr} ( [M^\dagger, M][B^\dagger, B]) \,, \ee where mesons ($M$) and
baryons ($B$) fall in the SU(3) representation $\ocho$, which is the
adjoint representation. The commutator indicates a $t-$channel
coupling to the $\ocho_a$ (antisymmetric) representation. 
Assuming the SU(8) spin--flavor symmetry, the mesons $M$ fall now in the
$\sesentaytres$ (adjoint representation) and the baryons $B$ are found
in the $\cientoveinte$, which is fully symmetric. The group reduction
lead to a total of four different $t-$channel SU(8) singlet couplings \cite{magas},
that can be used to construct $s$-wave meson-baryon interactions. However,
to ensure that the SU(8) amplitudes will reduce to those deduced from
the SU(3) WT Lagrangian in the
$(\ocho_\uno)$meson--$(\ocho_\dos)$baryon subspace (denoting the SU(3)
multiplets of dimensionality $\bf n$ and spin $J$ by ${\bf
n}_{\dos{\bf J}+\uno}$), we set all the couplings to be zero except for
\begin{equation}
{\cal L_{\rm WT}^{\rm SU(8)}}=
\left((M^\dagger\otimes M)_{\sesentaytres_a}\otimes
(B^\dagger\otimes B)_{\sesentaytres}\right)_{\uno}  \ ,
\label{eq:NOcoupl}
\end{equation}
which is the natural and unique SU(8) extension of the usual SU(3) WT
Lagrangian. To compute the matrix elements of the SU(8)
WT interaction,  ${\cal L_{\rm WT}^{\rm SU(8)}}$, we use quark model constructions of hadrons 
with field theoretical methods to express everything
in tensor representations as
described in  Appendix A of Ref.~\cite{magas}.
Since SU(8) spin-flavor symmetry is strongly broken in nature, we implement mass-breaking effects by adopting the physical hadron masses in the tree level interactions and in the evaluation of the
kinematical thresholds of different channels. Moreover, we consider the difference between
the weak non-charmed and charmed pseudoscalar and vector meson decay constants. Then, our tree level amplitudes read
\begin{eqnarray}
&&\hspace{-1cm}V^{IJSC}_{ab}(\sqrt{s}) = \nonumber \\ 
&&\hspace{-1cm}D^{IJSC}_{ab}
\frac{2\sqrt{s}-M_a-M_b}{4\,f_a f_b} \sqrt{\frac{E_a+M_a}{2M_a}}
\sqrt{\frac{E_b+M_b}{2M_b}} \,, \label{eq:vsu8break}
\end{eqnarray}
where  $IJSC$ are the meson-baryon isospin, total angular momentum, strangeness and charm quantum numbers, $M_a$ ($M_b$) and $E_a$ ($E_b$) the mass and
the CM energy, respectively, of the baryon in the $a$ ($b$) channel, $f_a$ ($f_b$) the weak decay constant of the meson in the $a$ ($b$) channel, and $D^{IJSC}$ a matrix of coefficients in the coupled-channel space \cite{magas}. 

\section{Non-strange single charmed baryon resonances}

With the kernel of the SU(8) WT meson-baryon interaction given in
Eq.~(\ref{eq:vsu8break}), we solve the coupled-channel on-shell
Bethe-Salpeter equation
\begin{eqnarray}
&&T^{IJSC}(\sqrt{s}) = \nonumber \\
&& \frac{1}{1-V^{IJSC}(\sqrt{s})\,G^{0\,(IJSC)}(\sqrt{s})}\,V^{IJSC}(\sqrt{s}) \ .
 \label{eq:scat-eq}
\end{eqnarray}
The loop function for each channel $a$,  $G^{0\,(IJSC)}_{a}(\sqrt{s})$, is divergent and is regularized 
by one-subtraction at the subtraction point $\mu^{ISC}$
\begin{eqnarray}
&&G^{0\,(IJSC)}_{a}(\sqrt{s}=\mu^{I}) = 0\, \nonumber \\
&&\left (\mu^{ISC}\right)^2=\alpha
\left(m_{\text{th}}^2+M^2_{\text{th}}\right)\ ,
\label{eq:sp}
\end{eqnarray}
where $m_{\text{th}}$ and $M_{\text{th}}$ are the meson and baryon
masses of the hadronic channel with lowest mass threshold for a fixed
$ISC$ and arbitrary $J$. The value of $\alpha=0.9698$ is adjusted to
reproduce the position of the well established $\Lambda_c(2595)$
resonance with $IJSC=(0,1/2,0,1)$.

The mass and widths of the dynamically generated baryon resonances are determined by the pole position, $z_R$, in the second Riemann sheet of the corresponding scattering amplitudes, namely $M_R={\rm Re}(z_R)$ and $\Gamma_R=-2 \ {\rm Im}(z_R)$. The coupling constants of each resonance to the various baryon-meson states are obtained from the residues by fitting the amplitudes to
\begin{equation}
T^{IJSC}_{ij}(z)=\frac{g_i e^{i \phi_i} g_j e^{i \phi_j}}{(z-z_R)} \ , \label{eq:pole}
\end{equation}
for complex energy values $z$ close to the pole, 
where the complex couplings are written in terms of the absolute value, $g_k$, and phase, $\phi_k$.
We will examine the $ij$-channel independent quantity
\begin{equation}
\label{obs} \tilde{T}^{IJSC}(z) \equiv \max_j\ \sum_i\,
|T^{IJSC}_{ij}(z)|\ ,
\end{equation}
which allows us to identify all the resonances within a given sector
at once. Among all dynamically generated baryon resonances of Ref.~\cite{magas}, we focus on  nonstrange ($S=0$) and singly charmed ($C=1$) baryon resonances in the $I=0$ and $I=1$ sectors that have or might have experimental confirmation \cite{yao} and compare to SU(4) results \cite{kolo,hofmann,angels-mizutani}. Note that the spin-parity of some resonances, such as the $\Sigma_c(2800)$ and $\Lambda_c(2940)$, are not determined yet experimentally.

\subsection{$I=0$, $J=1/2$}

In the $I=0$, $J=1/2$ sector we obtain the $\Lambda_c(2595)$ resonance with a width of $0.58$~MeV, that is smaller than the experimental one of $\Gamma=3.6^{+2.0}_{-1.3}$ \cite{yao}. This can be explained by the fact that in our calculation we have not included the three-body decay channel $\Lambda_c \pi \pi$, which represents one-third of the decay events \cite{yao}. We also observe a second resonance, $\Lambda_c(2610)$, very close to $\Lambda_c(2595)$, which seems to follow the double-pole pattern of the strange counterpart $\Lambda(1405)$ \cite{magas-angels-oset}. In our SU(8) scheme we find that the $\Lambda_c(2595)$ resonance is a predominantly $ND^*$ quasibound state as compared to SU(4) models where it emerges as $ND$ quasibound state.

Other dynamically generated resonances at higher energies are the $\Lambda_c(2822)$ and $\Lambda_c(2938)$ resonances. The first one is a $ \Xi_c K$ which does not correspond to the experimental $\Lambda_c(2880)$ \cite{yao} because of different spin-parity, but it is not incompatible with the $pD^0$ histogram \cite{aubert}. On the other hand, the $\Lambda_c(2938)$ cannot correspond to the experimental $\Lambda_c(2940)$ \cite{yao} because it does not couple to $ND$ states or not preferentially to $ND^*$, as discussed in \cite{he}.

\subsection{$I=1$, $J=1/2$}

In this sector we do not find any resonance between 2800 MeV and 3000 MeV which could correspond to the experimental $\Sigma_c(2800)$ \cite{yao}, because all the dynamically generated resonances in this energy region within the SU(8) scheme \cite{magas} fail to decay  primarily in $\Lambda_c \pi$ states although they couple quite significantly to $ND$ ones. Also our $\Sigma_c(3096)$ resonance cannot be identified with the experimental $\Sigma_c(2800)$ because this resonant state lies too high in mass to be moved to lower energies by changing the subtraction point. Likewise our $\Sigma_c(3035)$ resonance would be too narrow if it were moved to lower energies to make it compatible with the experimental $\Sigma_c(2800)$, even if the $\Lambda_c \pi \pi$ decay is allowed. 

Our $\Sigma_c(2974)$ resonance might correspond to the observed enhancement in the $I=1$ $D^+p$ histogram around 2860 MeV with a width of 10 MeV \cite{aubert}, if we lower its mass by changing the subtraction point as it will also reduce its width due to the closing of two out of three decaying channels.

\subsection{$I=0$, $J=3/2$}

The experimental $\Lambda_c(2625)$ and $\Lambda_c(2940)$ \cite{yao} might be seen in the $I=0$, $J=3/2$ sector. The $\Lambda_c(2625)$ with a width less than 2 MeV is the counterpart in the charm sector of the $\Lambda(1520)$. This resonance can be identified with our $\Lambda_c(2660)$ with a width of 38 MeV, which couples very strongly to the $\Sigma_c^* \pi$ channel. Changes in the subtraction point will move this resonance downwards and it will become substantially narrower as soon as it is relocated below the $\Sigma_c^* \pi$ threshold.
 On the other hand, the dynamically generated $\Lambda_c(2941)$ might be the experimental $\Lambda_c(2940)$ if we implement $p$-wave interactions in our SU(8) scheme in order to account for the $D^0p$ decay \cite{aubert}.

\subsection{$I=1$, $J=3/2$}

In this last sector we obtain the $\Sigma_c(2550)$ resonance, which couples strongly to $\Delta D$ and $\Delta D^*$ states. This state could be identified as the counterpart in the charm sector of the $\Sigma(1670)$, which decays primarily to $\Delta \bar K$. However, there is no experimental evidence so far. Moreover, our $\Sigma_c(2902)$ state in the $I=1$, $J=3/2$ sector of the SU(8) scheme might be identified with the experimental $\Sigma_c(2800)$ if this resonance could also be seen in $\Lambda_c \pi\pi$ states.

\section{Comparison with SU(4) models }

Compared to SU(4) TVME models \cite{hofmann,angels-mizutani}, the SU(8) scheme includes vector mesons in order to be consistent with heavy-quark symmetry.  Another essential difference lies in the fact that the transition amplitudes between states with heavy mesons go as the inverse of a heavy-meson decay constant for each heavy meson involved, whereas in the SU(4) models the decay constant is kept fixed for all transitions. As a result, we find that the SU(8) model reproduces all resonances generated in the SU(4) approaches that couple strongly to the channels consisting of a pseudoscalar octet meson and a charmed baryon. On the other hand, due to the different pattern of flavor symmetry breaking, resonances in the SU(4) models that couple strongly to baryon-meson states containing a charmed meson and an uncharmed baryon are not necessarily reproduced within our SU(8) approach. However, the enlarged model space in SU(8) due to heavy-quark symmetry compensates largely for the reduced attraction, generating the same resonances as in the SU(4) models but with a quite different composition. This is the case of the $\Lambda_c(2595)$ resonance in the $I=0$, $J=1/2$ sector. Within SU(4) models this resonance is dynamically generated mainly from $ND$ states. Instead, the SU(8) scheme interprets this resonance as being mainly a $ND^*$ quasibound state.

\section{Conclusions and Outlook}

In this work we study charmed baryon resonances within a coupled-channel unitary approach that implements heavy-quark symmetry. This is achieved by extending the $t$-channel vector-meson exchange SU(4) models to SU(8) spin-flavor symmetry and, then, implementing a somewhat different way of breaking the flavor symmetry through physical hadron masses and introducing the physical weak decay constants of the mesons involved in the transitions.

The SU(8) model generates a broad spectrum of baryon resonances with negative parity in all the isospin-spin sectors that one can form from an $s$-wave interaction between the mesons of the $0^-$, $1^-$ multiplets with the $1/2^+$, $3/2^+$ baryons. We focus in the $S=0$ and $C=1$ sector and on those resonances which already have or might have an experimental confirmation. In the $I=0$, $J=1/2$ we reproduce the experimental $\Lambda_c(2595)$, while in the $I=0$, $J=3/2$ we  assign the $\Lambda_c(2660)$ to the experimental $\Lambda_c(2625)$, which is the counterpart of the $\Lambda(1520)$. Similarly, in the $I=1$, $J=3/2$ we find the $\Sigma_c(2550)$, as the counterpart of $\Sigma(1670)$. In this sector, we might also identify the experimental $\Sigma_c(2800)$ with our $\Sigma_c(2902)$.  This broad spectrum of baryon resonances also includes the resonances that were generated in the previous SU(4) models. However, some of them have different nature, as in the case of the $\Lambda_c(2595)$. 

In order to make a more realiable comparison with experiments, future work implies the development of a more realistic model which contains three-body channels and higher-multipolarity interactions. We note that the experimental spectra show a limited amount of counts and, in order to disentagle new resonant structures from data, more statistics is definitely needed. The incorporation of medium modifications on those resonances \cite{laura} will also give us some insight into the nature of those states as well as the excitation mechanisms in the medium.  

\acknowledgments{L.T.  acknowledges support from the Rosalind Franklin Programme of the
 University of Groningen and the Helmholtz
 International Center for FAIR (LOEWE programme of State of Hesse, Germany). T.M. wishes to express his appreciation
for the support of his visit from Universitat de  Barcelona. This work is partly supported by the EU contract
FLAVIAnet MRTN-CT-2006-035482,
by the contract FIS2008-01143
from MEC (Spain) and
FEDER,  by the Generalitat de
Catalunya contract 2005SGR-00343,
and the Junta de Andaluc{\'\i}a
grants FQM225, FQM481 and P06-FQM-01735.
This research is part of the EU
Integrated Infrastructure Initiative Hadron Physics Project under
contract number RII3-CT-2004-506078.}

\end{multicols}

\begin{multicols}{2}

\end{multicols}


\begin{thebibliography}{90}

\vspace{3mm}

\bibitem{cleo}  M.~Artuso {\it et al.}  [CLEO Collaboration],
  Phys.\ Rev.\ Lett., 2001, {\bf 86}, 4479 

\bibitem{belle} R.~Mizuk {\it et al.}  [Belle Collaboration],
  Phys.\ Rev.\ Lett., 2005, {\bf 94}, 122002; 
  R.~Chistov {\it et al.}  [BELLE Collaboration],
  Phys.\ Rev.\ Lett., 2006, {\bf 97}, 162001.
 
\bibitem{babar}   B.~Aubert {\it et al.}  [BABAR Collaboration],
  Phys.\ Rev.\ Lett., 2006, {\bf 97}, 232001.

\bibitem{aubert} 
  B.~Aubert {\it et al.}  [BABAR Collaboration],
  Phys.\ Rev.\ Lett., 2007, {\bf 98}, 012001.

\bibitem{tolos-schaffne-mishra}  L.~Tolos, J.~Schaffner-Bielich and A.~Mishra,
  Phys.\ Rev.\  C, 2004,  {\bf 70}, 025203.


\bibitem{kolo}
  M.~F.~M.~Lutz and E.~E.~Kolomeitsev,
  Nucl.\ Phys.\  A, 2002, {\bf 700}, 193.


\bibitem{hofmann}
 J.~Hofmann and M.~F.~M.~Lutz,
  Nucl.\ Phys.\  A, 2005, {\bf 763}, 90.

\bibitem{hofmann2}
J.~Hofmann and M.~F.~M.~Lutz,
  Nucl.\ Phys.\  A, 2006, {\bf 776}, 17.

\bibitem{angels-mizutani}
  T.~Mizutani and A.~Ramos,
  Phys.\ Rev.\  C, 2006, {\bf 74}, 065201.

\bibitem{haide1}
  J.~Haidenbauer, G.~Krein, U.~G.~Meissner and A.~Sibirtsev,
  Eur.\ Phys.\ J.\  A, 2007, {\bf 33}, 107;
 J.~Haidenbauer, G.~Krein, U.~G.~Meissner and A.~Sibirtsev,
  Eur.\ Phys.\ J.\  A, 2008, {\bf 37}, 55.


\bibitem{Garcia-Recio:2005hy}
C.~Garc{\'\i}a-Recio, J.~Nieves and L.~L. Salcedo, Phys. Rev. D, 2006,
\textbf{74}, 034025.

\bibitem{GarciaRecio:2006wb}
  C.~Garcia-Recio, J.~Nieves and L.~L.~Salcedo,
  Phys.\ Rev.\  D, 2006, {\bf 74}, 036004.

\bibitem{Toki:2007ab} 
  H.~Toki, C.~Garcia-Recio and J.~Nieves,
  Phys.\ Rev.\  D, 2008, {\bf 77}, 034001.

\bibitem{magas}
C.~Garcia-Recio, V.~K.~Magas, T.~Mizutani, J.~Nieves, A.~Ramos, L.~L.~Salcedo and L.~Tolos,
  Phys.\ Rev.\  D, 2009, {\bf 79}, 054004.

\bibitem{yao}
  C.~Amsler {\it et al.}  [Particle Data Group],
  Phys.\ Lett.\  B, 2008, {\bf 667}, 1.

\bibitem{magas-angels-oset}  V.~K.~Magas, E.~Oset and A.~Ramos,
  Phys.\ Rev.\ Lett., 2005,  {\bf 95}, 052301.

\bibitem{he}  X.~G.~He, X.~Q.~Li, X.~Liu and X.~Q.~Zeng,
  Eur.\ Phys.\ J.\  C, 2007, {\bf 51}, 883.

\bibitem{laura} L. Tolos, C. Garcia-Recio and J. Nieves, arXiv:0905.4859 [nucl-th].

\end{thebibliography}
\end{document}